\documentclass{article}

\usepackage{PRIMEarxiv}

\usepackage[utf8]{inputenc} 
\usepackage[T1]{fontenc}    
\usepackage{hyperref}       
\usepackage{url}            
\usepackage{booktabs}       
\usepackage{amsfonts}       
\usepackage{nicefrac}       
\usepackage{microtype}      
\usepackage{amsmath}
\usepackage{fancyhdr}       
\usepackage{graphicx}       
\usepackage{multirow}%
\usepackage{makecell}%
\usepackage{float}
\graphicspath{{media/}}     

\title{Alleviating Seasickness through Brain–Computer Interface-based Attention Shift
}

\author{
  Xiaoyu Bao \thanks{The two authors contribute equally to this work.}\\
  School of Automation Science and Engineering \\
  South China University of Technology \\
  Guangzhou, China\\
  Research Center for Brain--Computer Interface\\
  Pazhou Laboratory\\
  Guangzhou, China\\
  \texttt{bkxcyu@outlook.com} \\
  \And
  Kailin Xu \textsuperscript{$*$} \\
  School of Automation Science and Engineering \\
  South China University of Technology \\
  Guangzhou, China\\
  Research Center for Brain--Computer Interface\\
  Pazhou Laboratory\\
  Guangzhou, China\\
  \texttt{kailinxu1998@163.com} \\
  \And
  Jiawei Zhu \\
  School of Automation Science and Engineering \\
  South China University of Technology \\
  Guangzhou, China\\
  Research Center for Brain--Computer Interface\\
  Pazhou Laboratory\\
  Guangzhou, China\\
  \texttt{zjw7342@163.com} \\
   \And
  Haiyun Huang \\
  School of Artificial Intelligence \\
  South China Normal University \\
  Foshan, China\\
  \texttt{huanghaiyun@m.scnu.edu.cn} \\
  \AND
  Kangning Li\\
  School of Psychology \\
  South China Normal University \\
  Guangzhou, China\\
  \texttt{15626460572@qq.com} \\
  \And
  Qiyun Huang \\
  Research Center for Brain--Computer Interface\\
  Pazhou Laboratory\\
  Guangzhou, China\\
  \texttt{huangqiyun@pazhoulab.cn} \\
  \And
  Yuanqing Li \thanks{Corresponding author} \\
  School of Automation Science and Engineering \\
  South China University of Technology \\
  Guangzhou, China\\
  Research Center for Brain--Computer Interface\\
  Pazhou Laboratory\\
  Guangzhou, China\\
  South China Brain--Computer Interface Technology Co., Ltd.\\
  Guangzhou, China\\
  \texttt{auyqli@scut.edu.cn} \\
}

\begin{document}
\maketitle

\begin{abstract}
Seasickness poses a widespread problem that adversely impacts both passenger comfort and the operational efficiency of maritime crews. Although attention shift has been proposed as a potential method to alleviate symptoms of motion sickness, its efficacy remains to be rigorously validated, especially in maritime environments. In this study, we develop an AI–driven brain–computer interface (BCI) to realize sustained and practical attention shift by incorporating tasks such as breath counting. Forty-three participants completed a real-world nautical experiment consisting of a real–feedback session, a resting session, and a pseudo–feedback session. Notably, 81.39\% of the participants reported that the BCI intervention was effective. EEG analysis revealed that the proposed system can effectively regulate motion sickness EEG signatures, such as an decrease in total band power, along with an increase in theta relative power and a decrease in beta relative power. Furthermore, an indicator of attentional focus, the theta/beta ratio, exhibited a significant reduction during the real–feedback session, providing further evidence to support the effectiveness of the BCI in shifting attention. Collectively, this study presents a novel nonpharmacological, portable, and effective approach for seasickness intervention, which has the potential to open up a brand–new application domain for BCIs.
\end{abstract}


\section{Introduction}\label{sec1}
The motions of a vessel at sea can elicit symptoms of motion sickness, commonly referred to as seasickness, which presents considerable challenges for both individuals working and traveling onboard such vessels. Seasickness not only leads to increased energy expenditure, but also exacerbates fatigue and drowsiness, particularly during prolonged maritime sailing and in the case of adverse sea conditions. Data collected from surveys conducted on various vessels underscore the substantial impact of seasickness. Reports indicate that approximately 64\% of inexperienced passengers experience seasickness during a 2--to--3--day cruise \cite{gahlinger_cabin_2006,krueger_controlling_2011}. Remarkably, in the case of rough sea conditions, the incidence of seasickness can reach a staggering 100\% \cite{kozarsky_prevention_1998}. Furthermore, seasickness poses a serious threat to the operational performance of military personnel. In amphibious naval exercises, the prevalence of seasickness can reach 62.5\%, and for naval medical personnel engaged in distant sea missions, the rate of seasickness can reach 72.35\% \cite{qi2021sea,visser2020patterns}. Therefore, the development of effective methods to address seasickness is a critical task.

As a subtype of motion sickness, seasickness is triggered primarily by a discrepancy between the sensory information perceived by the eyes/ears and the vestibular (inner ear) signals related to motion, leading to confusion in the brain and subsequent symptoms \cite{reason1978motion}. Pharmacological interventions have been used to alleviate symptoms by targeting the neural pathways involved in the motion response of the vestibular system. For example, medications such as dimenhydrinate, promethazine, cyclizine, and scopolamine work by inhibiting specific receptors in the brain that transmit signals related to motion and nausea. These medications have efficacy rates ranging from 62\% to 78\%  \cite{gordon2001effects, motamed2017comparison, wood1966clinical}. However, pharmacological interventions have limitations due to the notable side effects they cause, including headaches, xerostomia (dry mouth), fatigue, dizziness, and blurred vision  \cite{motamed2017comparison, schmal2013neuronal}. These side effects not only diminish individual comfort but may also impair cognitive functions, thereby negatively impacting the operational performance of maritime personnel. Furthermore, in maritime settings where medical supplies can be limited and difficult to procure, the reliance on these medications becomes problematic. Certain medications are subject to stringent restrictions in naval contexts owing to their associated high--risk profiles, further complicating their use \cite{powell-dunford2017}. Consequently, there is increasing interest in the exploration and development of nonpharmacological alternatives for managing seasickness, particularly interventions that are safe, accessible, easy to implement and not reliant on scarce medical resources.

Owing to its apparent simplicity and practicality, attention shift has been tentatively adopted as a potential method of alleviating motion sickness; however, the uncertainty in its efficacy in real-world scenarios remains to be addressed and resolved. This method involves shifting one’s attentional focus from the internal sensations of movement and nausea to external visual stimuli or engaging activities, thereby mitigating the discrepancy between visual and vestibular inputs and alleviating symptoms \cite{bang2023motion,wickens2014structure,venkatakrishnan2024}. Some attention shifting approaches have been proposed to alleviate motion sickness, such as listening to music \cite{sang2003behavioral,keshavarz2014}, regulating breathing patterns\cite{sang2003behavioral,russell2014use}, and performing mental arithmetic tasks \cite{bos2015}. However, the supporting evidence was primarily from simulated experiments conducted in laboratory settings within small groups of participants, lacking direct and rigorous validation in real--world environments.  Compared to other motion sickness scenarios, sea travel poses unique challenges, such as long voyage duration and intense rocking motions. The prolonged and continuous motion environment in maritime contexts, characterized by shaking, rolling, and pitching, can greatly impair the ability to shift and maintain attention on external visual cues or engage in activities \cite{bos2022,o1974motion}. Almost everyone is prone to seasickness during a long-term maritime journey \cite{kozarsky_prevention_1998}. Therefore, developing an effective and sustainable method for shifting attention to alleviate seasickness symptoms in maritime settings remains a highly formidable challenge.

To sustain attention shifts and alleviate seasickness symptoms during actual maritime navigation, we incorporate low cognitive load tasks derived from mindfulness meditation to shift attention, and leverage BCI to enhance their sustainability and practicality. In these tasks, individuals are directed to concentrate on the present moment or a specific object (e.g. breath) in a mindfulness manner \cite{brandmeyer2019neuroscience}, which could enhance attention engagement more effectively than simply breathing \cite{brandmeyer2019neuroscience}. However, low cognitive load tasks is easily disrupted by external distractions \cite{montero2021teachers}. The BCI establishes a direct link between the brain and external devices, enabling functions such as the capture and recognition of brain signals, real–time online assessment of brain states, and the delivery of recognition results to participants through closed–loop neurofeedback \cite{sitaram2017closed,wolpaw2002brain}. This approach helps participants achieve self–monitoring, which can enhance the effectiveness of tasks of mindfulness meditation and reduce the cost of learning, and provide a reliable and stable, endogenous focus of attention shift \cite{acabchuk2021measuring,stockman2020can}. Consequently, combining a BCI with low cognitive load tasks is though to assist participants in sustaining attention shifts in complex environments and holds significant potential for applications in scenarios such as alleviating seasickness.

We performed real--world maritime sailing experiments to evaluate the effectiveness of the BCI in reducing seasickness. A cohort of 43 participants engaged in three distinct sessions: a real--feedback mindfulness state (RMS), a pseudo--feedback mindfulness state (PMS), and a resting state (RS). The experimental outcomes confirmed the efficacy of the BCI in mitigating seasickness, as quantified by the MIsery SCale (MISC) score \cite{bos2005motion}. Moreover, EEG analysis demonstrated the system’s capability to effectively modulate EEG signatures associated with motion sickness. These EEG signatures include spectrum power across all frequency bands \cite{chen2010}, theta relative power \cite{chen2010,krugliakova2020changes,malik2015eeg,wood1994habituation,wood1991effect,wu1992eeg} and beta relative power \cite{naqvi2015,kim2005characteristic}. Additionally, a well--established measure of attentional focus, the theta/beta ratio, showed a significant reduction during the real--feedback mindfulness session, providing further substantiation for the BCI's effectiveness in redirecting attention. Furthermore, we offer mechanistic insights into how the BCI alleviate seasickness, aligning with theories of attention shifting and sensory conflict.

\section{Results}\label{result}


We use a BCI headband to capture and amplify prefrontal EEG signals, which are wirelessly transmitted to a computing device for attention scores assessment. participants receive feedback through attention scores and 2D/3D audiovisual feedback, offering real--time insight into their mindfulness state to improve focus and engagement in the tasks. Forty--three individuals participated in real--world maritime sailing experiments, which comprised three sessions: RMS, RS and PMS. This section presents the behavior results of experiments exploring the use of BCI techniques for alleviating seasickness, along with the corresponding EEG signatures observed.

\subsection{Effective alleviation of BCI on seasickness symptoms}\label{result_1}
\begin{figure*}[htbp]
    \centering
    \includegraphics[width=11.4cm]{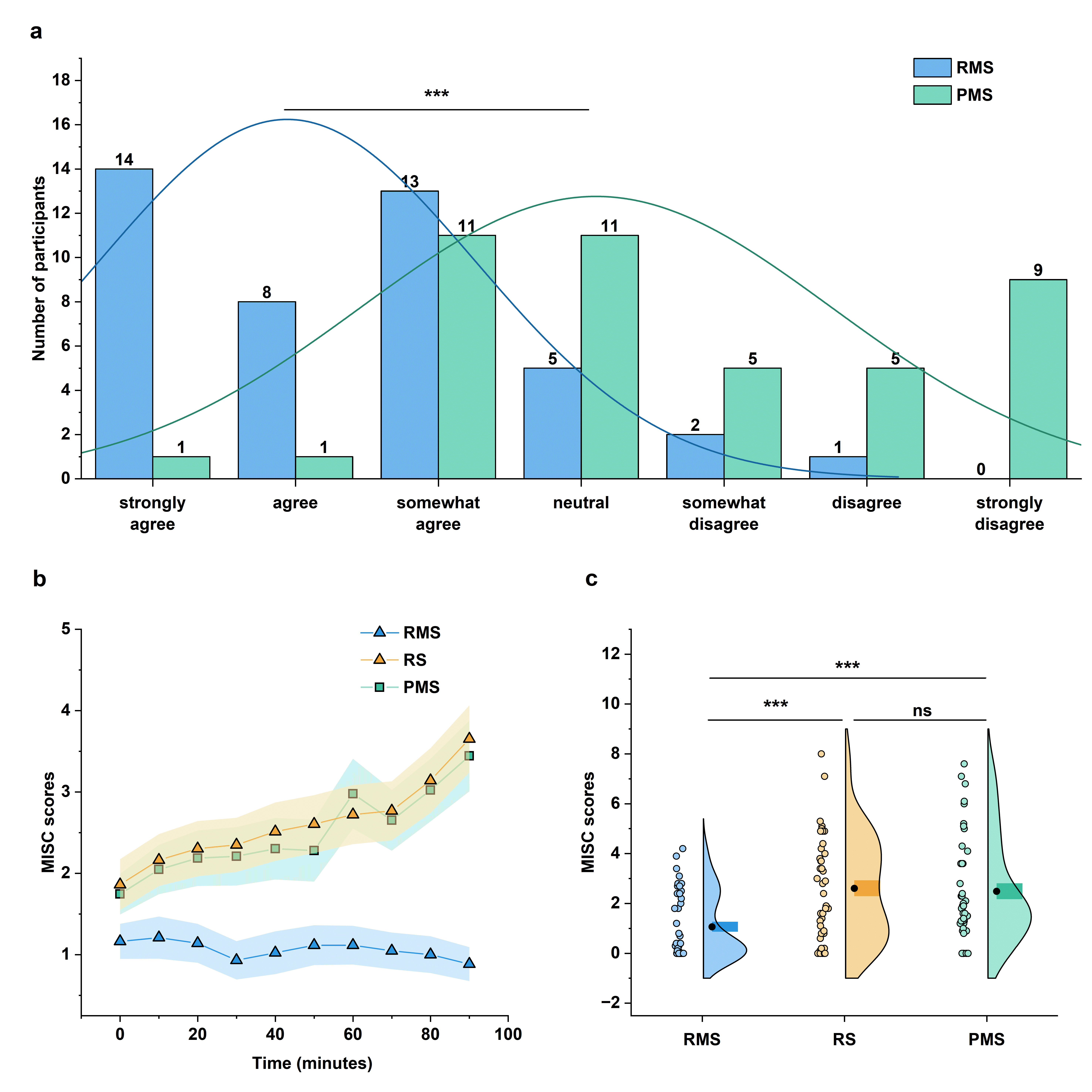}
    \caption{Behavioral results of the real--feedback mindfulness state (RMS), resting state (RS) and pseudofeedback mindfulness state (PMS).
   \textbf{a}, Assessment of participants regarding the efficacy of seasickness relief during the RMS and PMS. A 7--point Likert scale ranging from strongly agree (1 point) to strongly disagree (7 points) was utilized in the evaluation. The results in the RMS (mean = 2.442, s.e.m. = 0.201) and PMS (mean = 4.605, s.e.m. = 0.252) revealed a significant difference between the two states (two--tailed, paired $t(42)=-6.816, P<0.001$, Cohen's $d=-1.039$). \textbf{b}, Mean MISC scores of all participants in the RMS, RS and PMS. The shaded areas represent the standard error of the mean. \textbf{c}, MISC scores in the RMS, RS and PMS. Each data point represents the time--averaged MISC score of one participant. A significant difference (two--tailed, paired $t(42)=-4.785, P<0.001$, Cohen's $d=-0.730$, FDR--corrected) was observed in the mean MISC scores between the RMS (mean = 1.063, s.e.m. = 0.200) and the RS (mean = 2.607, s.e.m. = 0.316). A significant difference (two--tailed, paired $t(42)=-5.147, P<0.001$, Cohen's $d=-0.785$, FDR--corrected) in the mean MISC scores between the RMS (mean = 1.063, s.e.m. = 0.200) and the PMS (mean = 2.486, s.e.m. = 0.318) was also observed. No significant difference in the mean MISC score was observed between the RS and PMS. The boxes within the half--violin plot denote the standard error of the mean, while the central black dots denote the means. ***: $P < 0.001$, ns: not significant.}\label{res_behavi_e1e2}
    \label{fig:enter-label}
\end{figure*}

After conducting real--world maritime sailing experiments involving the RMS, RS and PMS, the participants evaluated the effectiveness of the BCI in mitigating seasickness in both the RMS and PMS using the 7--point Likert scale  (Figure \ref{res_behavi_e1e2}a). Most of the participants (35 out of 43, 81.39\%) reported that the BCI effectively reduced seasickness in the RMS . On the contrary, in the PMS, the majority of the participants (n=30, 69.77\%) reported no relief from seasickness  . The comparisons of the 7--point Likert scale results between the RMS and PMS revealed a significant difference  (two--tailed, paired $t(42)=-6.816, P<0.001$, Cohen's $d=-1.039$).

We then examined the temporal effects of the BCI system in alleviating seasickness symptoms that accessed by MISC in three 90--minute sessions (Figure \ref{res_behavi_e1e2}b). A significant interaction effect was identified between state and time   ($F(18) =4.205, P < 0.001$). The analysis of main effects demonstrated that the MISC score was significantly influenced by both time  ($F(9) = 10.593, P < 0.001$) and state ($F(2) = 9.168, P < 0.001$). Post hoc analyses revealed that initially, at 0 minutes, no significant difference in the MISC scores was observed between the RMS and RS (two--tailed, paired $t(42) = -1.934, P = 0.137$, FDR--corrected). However, starting from the $20^{\text{th}}$ minute, the MISC scores in the RMS were significantly lower than those in the RS were (two--tailed, paired t--test, $P < 0.05$ every 10 minutes, FDR--corrected). Therefore, the increase in the MISC score over time was significantly mitigated by using the BCI. 

The MISC scores were subsequently averaged within the RMS, the RS and the PMS. As presented in Figure \ref{res_behavi_e1e2}c, the averaged MISC score in the RMS was significantly lower than those in the RS (two--tailed, paired $t(42)=-4.128, P<0.001$, Cohen's $d=-0.044$, FDR--corrected) and PMS (two--tailed, paired $t(42)=-3.785, P<0.001$, Cohen's $d=-0.0448$, FDR--corrected). These results demonstrate the efficacy of the BCI in alleviating seasickness symptoms during real--world maritime sailing experiments. 

\subsection{Superior performance of BCI for individuals experiencing severe seasickness}

\begin{figure*}[htbp]
    \centering
    \includegraphics[width=11.4cm]{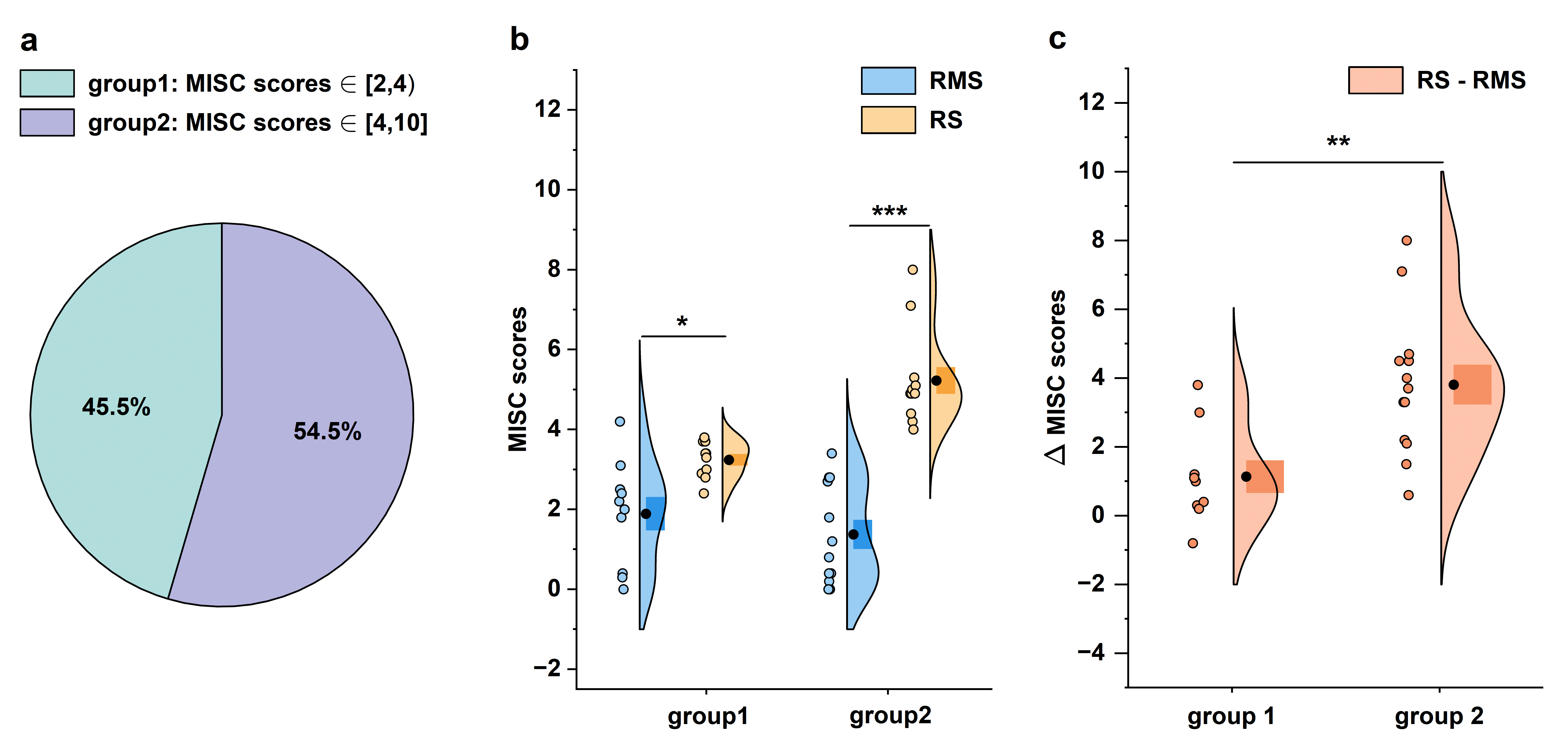}
    \caption{BCI is more effective for individuals experiencing severe seasickness.
    \textbf{a}, Proportion of participants with mild or severe seasickness classified by the time--averaged MISC score in the resting state (RS). Group 1 includes participants with mild seasickness symptoms (MISC score (RS) $\in [2,4)$, $n=10$), and group 2 includes participants with severe seasickness symptoms (MISC score (RS) $\in [4,10]$, $n=12$). 
    \textbf{b}, MISC scores of participants in group 1 and group 2 in the RS and real--feedback mindfulness state (RMS). Each point represents the time--averaged MISC score of one participant. In group 1, a significant difference (two--tailed, paired $t(9)=-2.816, P<0.05$, Cohen's $d=-1.360$, FDR--corrected) was observed between the mean score in the RMS (mean = 1.89, s.e.m. = 0.420) and RS (mean = 3.240, s.e.m. = 0.143). 
    In group 2, a significant difference in the MISC score (two--tailed, paired $t(11)=-6.160, P<0.001$, Cohen's $d=-3.169$, FDR--corrected) was also found between the RMS (mean = 1.375, s.e.m. = 0.364) and RS (mean = 5.225, s.e.m. = 0.336).
    \textbf{c}, Relative MISC scores of group 1 and group 2 participants in the RMS, with the score in the RS used as a baseline. Welch's t--test showed a significant difference (two--tailed, $t(19.501)= -3.174, P<0.01$, Cohen's $d=-1.315$) in the relative MISC scores between group 1 (mean = 1.35, s.e.m. = 0.479) and group 2 participants (mean = 3.85, s.e.m. = 0.625), with group 2 participants showing greater alleviation of seasickness symptoms compared with those in the RS. 
    The boxes within the half--violin plot denote the standard error of the mean, while the central black dots signify the means. ***: $P < 0.001$, **: $P < 0.01$, *: $P < 0.05$, ns: not significant.}\label{res_relief_e1e2}
\end{figure*}

We also observed an association between the severity of seasickness in the RMS, RS, and PMS. After excluding participants without clear seasickness symptoms (i.e., MISC score $<$ 2, $n = 21$) during the RS, the remaining participants were stratified into two groups: individuals with an MISC score greater than 2 but less than 4 were included in group 1 (10 out of 22, 45.5\%), and individuals with an MISC score of at least 4 were included in group 2 (12 out of 22, 54.5\%), as shown in Figure \ref{res_relief_e1e2}a.
Figure \ref{res_relief_e1e2}b shows that for group 1, there was a slight but statistically significant difference between the MISC scores in the RS and RMS (two--tailed, paired $t(9)=-3.041, P<0.05$, Cohen's $d=-1.360$, FDR--corrected). In contrast, in group 2, the MISC score in the RMS was significantly lower than that in the RS (two--tailed, paired $t(11)=-7.763, P<0.001$, Cohen's $d=-3.169$, FDR--corrected).
We subsequently assessed the variation in the MISC scores between the RMS and RS, as illustrated in Figure \ref{res_relief_e1e2}c. A significant difference (two--tailed, $t(19.501)= -3.174, P<0.01$, Cohen's $d=-1.315$) in the relative MISC scores was found between the participants in group 1 (mean = 1.35, s.e.m. = 0.479) and group 2 (mean = 3.85, s.e.m. = 0.625), with individuals in group 2 showing greater alleviation of seasickness symptoms compared with those in the RS. These results suggest that the BCI is particularly beneficial for individuals with a serve symptom of seasickness. 

\subsection{Modulation of potential EEG signatures of seasickness}

\begin{figure*}[htbp]
\centering
\includegraphics[width=11.4cm]{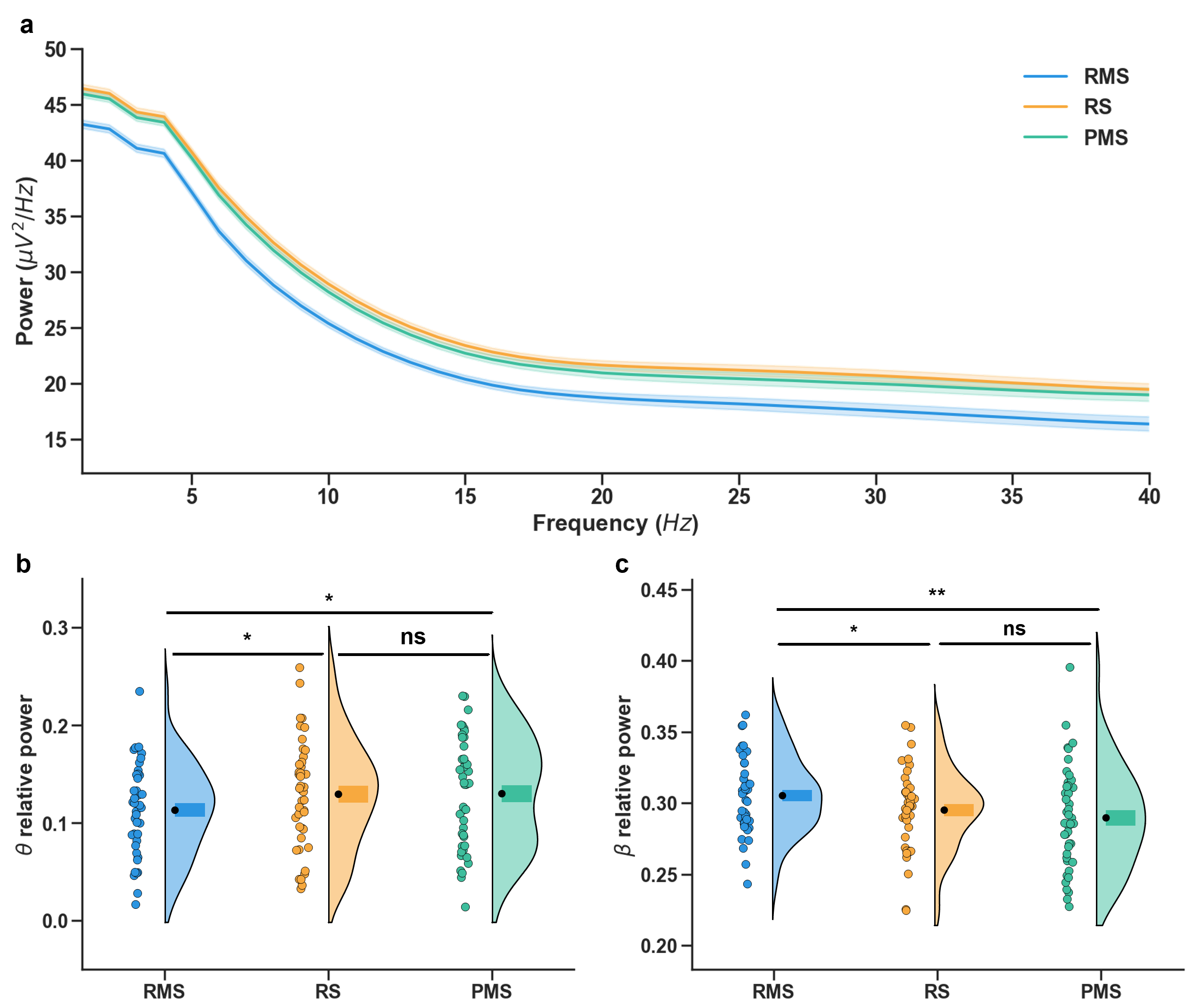}
\caption{EEG signature of seasickness. \textbf{a}, Power spectrum density curves in the three states (real--feedback mindfulness state (RMS), pseudofeedback mindfulness state (PMS), and resting state (RS)). The spectrum across the entire frequency band (1--40 Hz) was significantly lower in the RMS than in the RS and PMS (cluster--level statistical permutation test with a frequency resolution of 1 Hz, each frequency point $P<0.001$) \textbf{b}, The theta relative powers in the RMS, RS and PMS. The theta relative power in the RMS (mean = 0.11, s.e.m. = 0.007) was significantly lower than that in the RS (mean = 0.13, s.e.m. = 0.008) (two--tailed, paired $t(42) = -2.504, P < 0.05$, Cohen's $d = -0.382$, FDR--corrected) and PMS (mean = 0.13, s.e.m. = 0.009) (two--tailed, paired $t(42) = -2.377, P < 0.05$, Cohen's $d = -0.362$, FDR--corrected). \textbf{c}, The theta relative powers in the RMS, RS and PMS. The theta relative power was significantly greater in the RMS than in the RS (two--tailed, paired $t(42) = 2.772, P < 0.05$, Cohen's $d = 0.423$, FDR--corrected) and PMS (two--tailed, paired $t(42) = 3.383, P < 0.01$, Cohen's $d = 0.516$, FDR--corrected). The boxes and black dots within the half--violin plot represent the standard errors and means, respectively. *: $P < 0.05$, **: $P < 0.01$, ns: not significant. }\label{res_eeg_beta}
\end{figure*}

This section evaluates the neurophysiological changes induced by BCI by analyzing several potential EEG indicators of motion sickness as identified in prior research.  As shown in Figure \ref{res_eeg_beta}a, comprehensive power spectrum analysis was performed across the RMS, RS, and PMS, revealing that the spectrum was significantly lower in the RMS than in the RS and PMS across the entire frequency range (1–40 Hz) (cluster--level statistical permutation test with a frequency resolution of 1 Hz, each frequency point  $P<0.001$).

For relative power, we discovered that the prefrontal theta relative power, was significantly lower in the RMS than in the RS (two--tailed, paired $t(42) = -2.504, P < 0.05$, Cohen's $d = -0.382$, FDR--corrected) and PMS (two--tailed, paired $t(42) = -2.377, P < 0.05$, Cohen's $d = -0.362$, FDR--corrected), as shown in Figure \ref{res_eeg_beta}b. In addition, the prefrontal beta relative power was significantly greater in the RMS than in the RS (two--tailed, paired $t(42) = 2.772, P < 0.05$, Cohen's $d = 0.423$, FDR--corrected) and PMS (two--tailed, paired $t(42) = 3.383, P < 0.01$, Cohen's $d = 0.516$, FDR--corrected), as shown in  Figure \ref{res_eeg_beta}c. These results demonstrate that the BCI can effectively modulate the EEG signatures of seasickness. Furthermore, in the RMS, the prefrontal theta relative power negatively correlated with the beta relative power ($r=-0.712$, $P<0.001$). This correlation was also evident in the RS ($r=-0.719$, $P<0.001$) and in the PMS ($r=-0.859$, $P<0.001$), indicating a potential association in the underlying mechanism of these two signatures.

\subsection{Facilitative effect of BCI on attention shift}

\begin{figure*}[htbp]
\centering
\includegraphics[width=11.4cm]{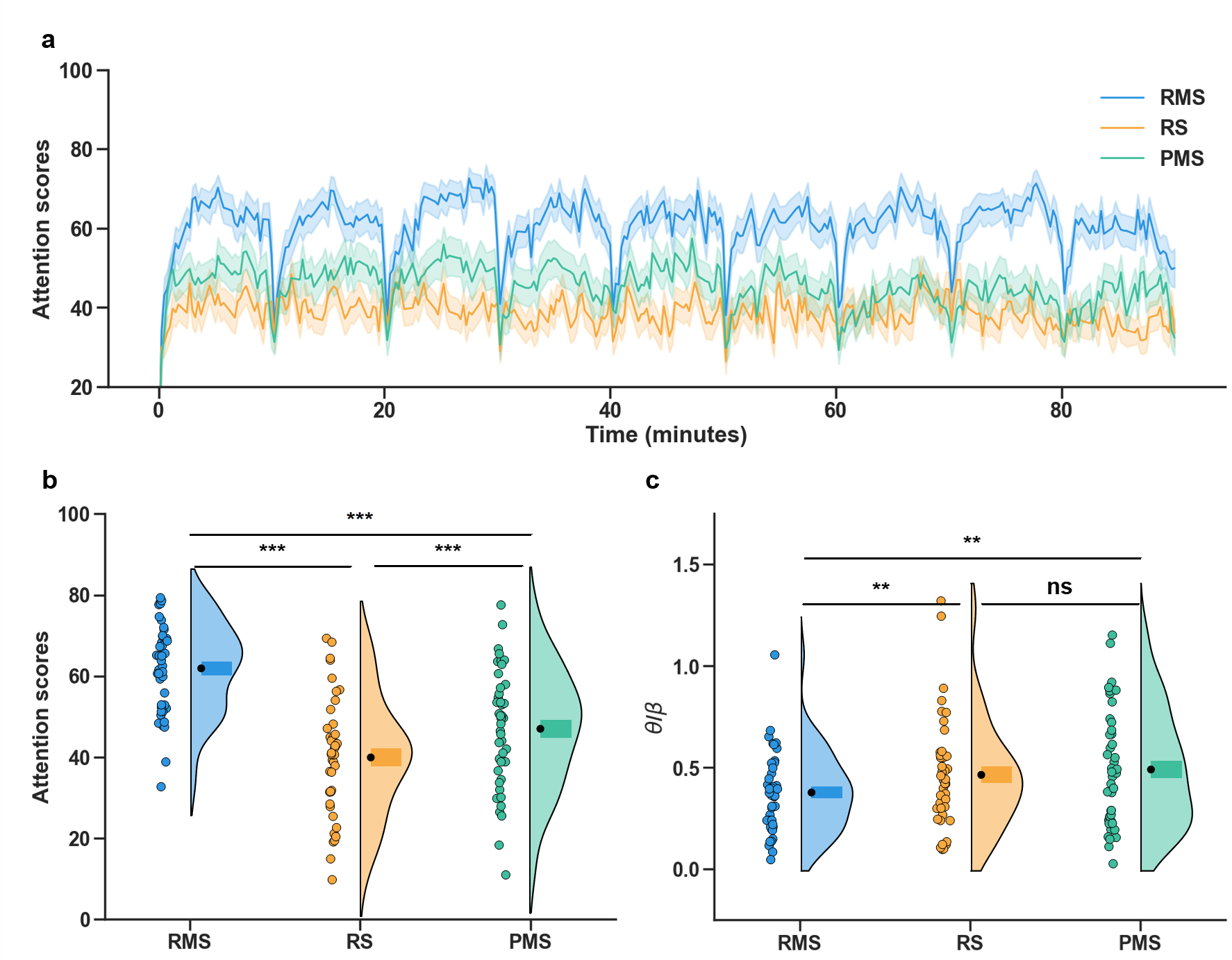}
\caption{attention scores and theta/beta power ratios related to attention. \textbf{a}, Average attention scores among subjects in the real--feedback mindfulness state (RMS), resting state (RS) and pseudofeedback mindfulness state (PMS). The shaded region represents the standard error of the mean. \textbf{b}, The average attention scores over time. The attention score in the RMS was significantly higher than that in the RS  (two--tailed, paired $t(42) = 10.671, P < 0.001$, Cohen's $d = 1.627$, FDR--corrected) and PMS (two--tailed, paired $t(42) = 7.146, P < 0.001$, Cohen's $d = 1.090$, FDR--corrected). The attention score in the PMS was also significantly higher than that in the RS (two--tailed, paired $t(42) = 3.914, P < 0.001$, Cohen's $d = 0.597$, FDR--corrected).  \textbf{c}, The theta/beta power ratio was significantly lower in the RMS than in the RS (two--tailed, paired $t(42) = -2.876, P < 0.01$, Cohen's $d = -0.439$, FDR--corrected) and PMS (two--tailed, paired $t(42) = -3.386, P < 0.01$, Cohen's $d = -0.516$, FDR--corrected). The boxes and black dots within the half--violin plot represent the standard errors and means, respectively.  **: $P < 0.01$, ***: $P < 0.001$, ns: not significant.
}\label{res_eeg_attention}
\end{figure*}

To evaluate attentional engagement using BCI, we analyzed both attention scores and the theta/beta power ratio, an established EEG marker of attention \cite{kim2022effect,katyal2021alpha}. As shown in Figure \ref{res_eeg_attention}a, the attention scores for participants in the RMS indicate notable decreases at the initial moment and at the beginning of every 10--minute interval, corresponding to the onset of meditation on the basis of a changed audiovisual scene. As shown in Figure \ref{res_eeg_attention}b, the attention score in the RMS was significantly higher than that in the RS (two--tailed, paired $t(42) = 10.671, P < 0.001$, Cohen's $d = 1.627$, FDR--corrected) and PMS (two--tailed, paired $t(42) = 7.146, P < 0.001$, Cohen's $d = 1.090$, FDR--corrected). The attention score in the PMS was also significantly higher than that in the RS (two--tailed, paired $t(42) = 3.914, P < 0.001$, Cohen's $d = 0.597$, FDR--corrected). These results suggest that the BCI effectively facilitated sustained attention for participants during tasks.

Figure \ref{res_eeg_attention}c presents EEG analysis results that corroborate these conclusions. The theta/beta ratio was significantly   lower in the RMS than in both the RS and PMS (RMS vs. RS: two--tailed, paired $t(42) = -2.876, P < 0.01$, Cohen's $d = -0.439$; RMS vs. PMS: two--tailed, paired $t(42) = -3.386, P < 0.01$, Cohen's $d = -0.516$; FDR--corrected). This decrease in the theta/beta ratio, which is known to be associated with increased attentional processing \cite{putman2014eeg,ogrim2012quantitative}, provides compelling neurophysiological support that participants were actively engaged in BCI--based tasks during the maritime sailing experiments.

\section{Materials and Methods}\label{sec11}

\subsection{Components and Functionality of the BCI System}
The proposed BCI system consists of three components: a BCI headband intended to acquire EEG signals, a computing device (e.g., mobile phone, tablet PC, or PC/laptop) to process EEG signals in real--time, and a software application designed to implement audiovisual neurofeedback. To carry out the experiment in real world maritime settings, tablet PCs (Lenovo® Xiaoxin Pad Pro 12.7”, Android® operating system, Snapdragon® 870 SoC, 2944×1840 screen resolution of 2944$\times$1840, screen refresh rate of 144 Hz) along with headphones were used.

The BCI headband is a single--channel EEG device (iHNNK, Inc.) characterized by a sampling rate of 250 Hz, with electrodes positioned above the right frontal lobe, and reference and ground electrodes situated at the temples. These electrodes are constructed from hydrogel material, which does not leave any residue on the participant's scalp, thus minimizing the preparation and cleaning time relative to the application of conventional conductive gels. The headband is portable, incorporating a circuit board for signal amplification, a Bluetooth module for wireless EEG data transmission, and operates with a built--in rechargeable battery.

The tablet PC can receive EEG signals transmitted from the BCI headband. The application installed on the tablet PC is capable of calculating attention scores (0 to 100) in real--time using EEG signals. These scores indicate the degree of concentration of the participant, where higher scores indicate a deeper sense of relaxation while maintaining inner focus.

The application incorporated a series of audiovisual scenes developed using Rive (rive.app), which serves as a framework for constructing animations characterized by sophisticated interactivity and state–driven animation paradigms. Each audiovisual scene was designed by a certified meditation instructor and came with meditation guidelines. These guidelines were formulated to facilitate participants' engagement in practices such as breath counting, body scanning, and auditory focus, executed in a mindfulness-oriented manner. Besides, audiovisual scenes adapt to attention scores, offering real–-time audiovisual feedback to participants. Specifically, higher scores improve the scenes’ clarity and aesthetics, as well as the volume of background sounds, whereas lower scores make the scenes vaguer and lessen the background sounds, with the volume of instructions remaining constant. For instance, within the ``campfire meditation" scene, participants are directed to engage in breath counting. A higher score leads to a more vibrant campfire and louder sounds, while a lower score reduces both. Participants are instructed to amplify the campfire’s flame intensity and sound volume by concentrating on breath counting. Through these tasks, participants can easily assess their attention state or recognize mind–wandering through audiovisual feedback, helping them to return their focus to the tasks. In general, participants would note that this kind of attention redirection tasks became more manageable and enjoyable with the BCI system.

\subsection{Real--time attention score assessment}

\begin{figure*}[tbhp]
\centering
\includegraphics[width=11.4cm]{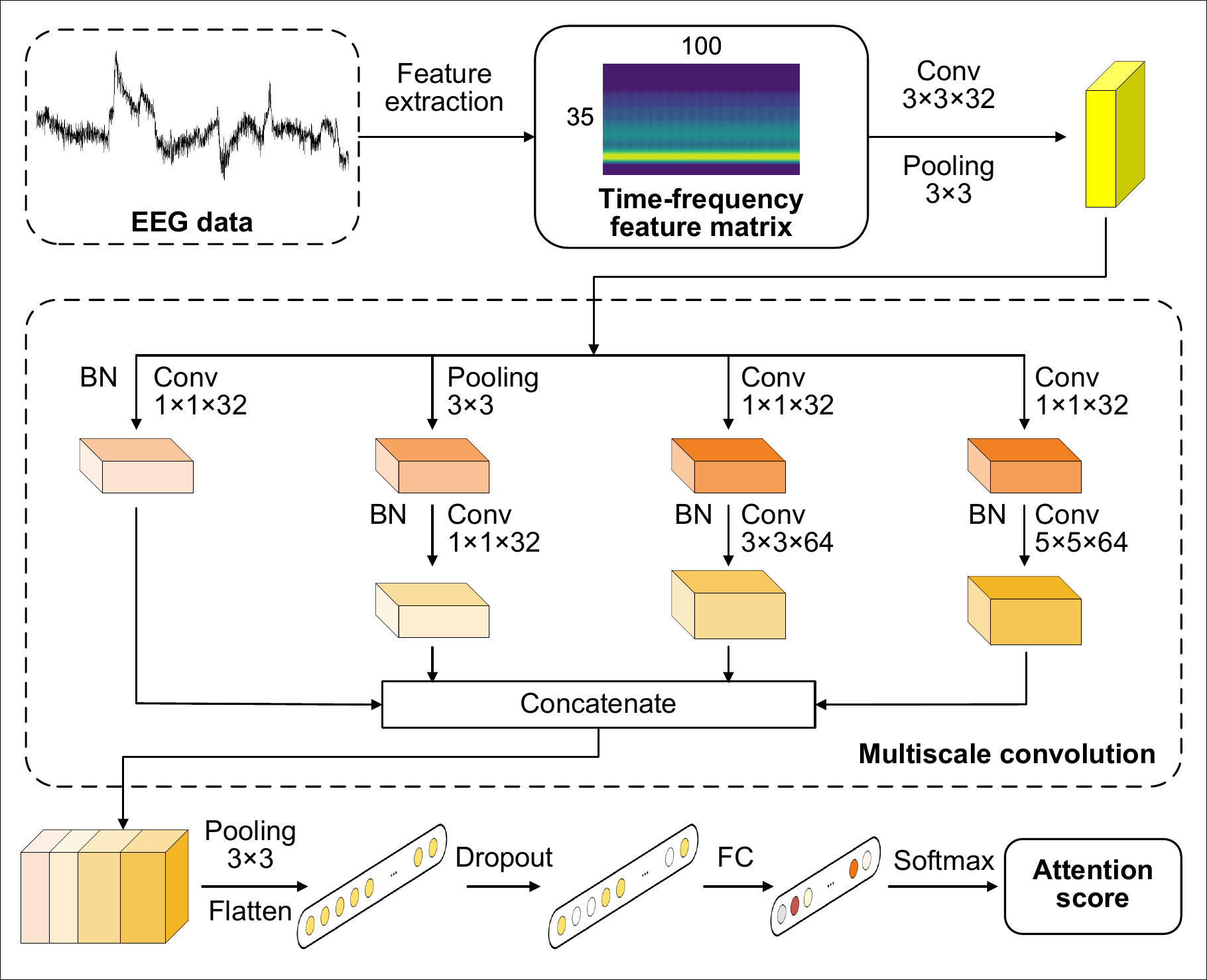}
\caption{Framework of the cross--subject CNN model used for attention score assessment. Conv, Pooling, BN and FC denote the convolution layer, max pooling layer, batch normalization layer and fully connected layer, respectively.}\label{method_cnn}
\end{figure*}

A cross--subject convolution neural network (CNN) model was implemented in the application to assess the attention scores of the participants in real--time. As illustrated in Figure \ref{method_cnn}, an EEG signal was extracted each second with a 10--second sliding window in real--time. A set of 35 bandpass filters, ranging from 0.1 Hz to 70.1 Hz in 2 Hz increments, was used to process each 10--second EEG signal. Then, for a specific EEG epoch, the time--frequency features were calculated using the Hilbert transformation \cite{feldman2011hilbert}. Afterward, the extracted time--frequency feature matrix was z score normalized and fed into the CNN model. 

Figure \ref{method_cnn} illustrates the CNN that uses the time–frequency EEG feature matrix as input. The first convolution layer starts by processing the input feature matrix with 32 kernels sized $3 \times 3$ and a stride of 1. Next, there is a max pooling layer with a pool size of $3 \times 3$ and a stride of 2. The features that result are fed into the multiscale convolution module, which has 4 branches with different receptive fields. Padding is used in the convolution and max--pooling layers to maintain feature dimensions. The features from each branch are merged and then directed to a max pooling layer with a pool size of $3 \times 3$ and a stride of 2, after which they are flattened into a vector. This vector passes through a dropout layer with a drop rate of 0.5 before entering a fully connected layer with a 100--size kernel. Ultimately, the classification is performed by a Softmax layer, which outputs a attention score, indicating the likelihood of the mindfulness state. All layers, except the output layer, use the rectified linear unit (ReLU) as the activation function.

The parameters of the CNN model were initialized using the Xavier initialization method and then updated through the Adam optimization algorithm. The model was trained in a Python environment on an NVIDIA GeForce GPU using an EEG dataset prepared in our previous research. The EEG dataset consists of EEG recordings taken during both focused--attention mindfulness meditation and rest states. Consequently, a cross--subject CNN model with optimized parameters was developed.

\subsection{Questionnaires}
At the end of the experiment, the participants evaluated the effectiveness of the proposed BCI system in mitigating seasickness using a 7--point Likert scale, which included strongly agree, agree, somewhat agree, neutral, somewhat disagree, disagree and strongly disagree. During the experiments, the participants evaluated their levels of seasickness using the Chinese version of the MISC table, as detailed in Table \ref{tab1}\cite{bos2005motion}. The scale, which ranges from 0 to 10, categorizes symptoms into five levels. A score of 0 indicates no problems, a score of 1 denotes discomfort without particular symptoms, scores from 2 to 5 correspond to symptoms with four increasing severity levels (vague, little, rather, and severe) without nausea, scores from 6 to 9 indicate symptoms with four increasing severity levels with nausea, and a score of 10 signifies vomiting. 

\begin{table}[h]
\caption{The MISC values used to measure 
seasickness severity}\label{tab1}
\centering
\begin{tabular}{ccc}
\toprule%
Symptoms     &     & MISC  \\
\midrule
No problems  &   & 0\\
\midrule
\multirow{2}{*}{\makecell{Uneasiness \\(no typical symptoms)}}  &   & 1\\
& & \\
\midrule
\multirow{5}{*}{\makecell{Dizziness, warmth,\\ headache,stomach,\\ awareness,sweating,\\and other symptoms, \\but no nausea}}
  & Vague  & 2\\
  & Slight  & 3\\
  & Fairly & 4\\
  & Severe  & 5\\
  & &\\
\midrule
\multirow{4}{*}{Nausea}
  & Slight  & 6\\
  & Fairly  & 7\\
  & Severe  & 8\\
  & Retching & 9\\
\midrule
 Vomiting &   & 10\\
\bottomrule
\end{tabular}
\end{table}
\subsection{Experiments}

Participants were recruited through online advertisements and word--of--mouth referrals. A total of 43 individuals signed up for the experiment. The mean age of the participants was 26.5 years, ranging from 22 to 34 years. The participants reported that they were in good health, possessing normal vision and auditory ability, without any history of neurological, cardiovascular, musculoskeletal, or vestibular disorders, and would not taking medications during the study. Participants retained the right to withdraw from the study at any time and gave informed consent according to protocols approved by the Committee for the Second Affiliated Hospital of the South China University of Technology (reference number: 2024--111--01).

The experiment included three sessions, an RMS, a PMS, and an RS, each lasting 90 minutes. The sequence of the sessions was randomized among participants, with an interval of at least 20 minutes between the sessions. The experiment started approximately an hour after the subject began experiencing discomfort on the research vessel. During each session, participants recorded their MISC scores by selecting options on the tablet PC every 10 minutes, and prefrontal EEG signals were collected using the BCI headband. 
In the RS, participants remained inactive. 
In the RMS, participants practiced BCI--assisted mindfulness meditation. 
During the PMS, participants engaged in pseudo--feedback mindfulness meditation with randomly changing audiovisual scenes. 
The attention scores of each state were automatically recorded by the application for analysis. After the three sessions, the participants evaluated the effectiveness of the mindfulness interventions of RMS and PMS in alleviating motion sickness on the 7--point Likert scale. 

The research vessel was provided by the Chinese Academy of Fisheries Sciences, weighs 399 tons, and was designed to travel at a speed of 12.5 knots. The experiment was carried out in the sea region of the Pearl River Delta, Guangdong, China, which spans approximately 80 nautical miles, and the entire experiment lasted approximately 10 hours. This procedure was repeated three times; each day, the wind speed was approximately 6 m / s and the wave height was approximately 1 m.

The participants were familiarized with the BCI system before the experiments began. The duration of this process varied according to the level of expertise of each participant. The participants were advised to focus on their breathing through deep breathing exercises and silent breath counting, which are essential meditation techniques. Typically, participants became familiar with the system in 1 to 3 practice sessions that lasted 10 to 20 minutes each.

\subsection{Behavioral data and EEG signature analysis}
The proportion of participants who selected positive opinions (i.e., strongly agree, agree, and somewhat agree) on the 7--point Likert scale was quantified. These positive opinions were interpreted as an endorsement of the effectiveness of the BCI in mitigating seasickness. The average MISC was accessed by averaging the MISC scores within three the states: RMS, PMS, and RS. The relative MISC scores for the mindfulness state were calculated by subtracting the scores of the RS.  

For EEG signature analysis, the raw EEG data was initially subjected to bandpass filtering with an infinite impulse response (IIR) filter, setting the frequency range between 1 Hz and 45 Hz. Post--filtering, the data was partitioned into 60--second epochs. Any epoch with an amplitude exceeding 300 $\mu V$ or falling below 10 $\mu V$ were discarded. The power spectral density of each EEG epoch was determined using the Welch method. This enabled the calculation of the power values for the frequency bands: theta (4 -- 8 Hz), alpha (8 -- 13 Hz), beta (13 -- 30 Hz), and the total frequency band (1 -- 40 Hz). These power values were then averaged for each state. Subsequently, the theta/beta ratios were determined. The relative power, which represents the contribution of each band to the overall EEG, was defined as the ratio of the power of a specific band to the power of the total frequency band.

\subsection{Statistical information}
The effects of state (RMS, PMS and RS) and time (intervals of 10 minutes) on the MISC scores were analyzed through repeated--measures ANOVA. Two--way ANOVA was employed to evaluate the impact of state (RMS, PMS and RS) on the average MISC scores, prefrontal theta power, prefrontal beta power, and theta/beta power ratio. Significant differences ($P<0.05$) were further examined with two--tailed paired t--tests with Benjamini--Hochberg false discovery rate (FDR) correction \cite{benjamini1995controlling}. Cohen’s  was utilized to calculate the effect sizes for for the significant results of the t--tests. To assess the coherence between theta relative power and beta relative power, Pearson’s correlation analysis was employed. All the statistical analyses were carried out using IBM SPSS 27.0.

\section{Discussion and conclusion}\label{discussion}
This study presents a novel nonpharmacological approach method for alleviating seasickness. It utilizes an AI--powered wearable BCI system specifically crafted to efficiently shift the participant's attention from seasickness towards tasks. In real--world maritime sailing experiments, the majority of participants reported that the BCI significantly alleviated their seasickness symptoms. Notably, the BCI proved particularly effective for individuals suffering from severe seasickness, as evidenced by a correlation between the symptom severity and the degree of relief. Moreover, the EEG signatures related to seasickness was modulated by the use of the BCI, offering subjective neurophysiological evidence supporting its efficacy in mitigating seasickness symptoms. The theta/beta ratio, an indicator of attentional focus, was also found to decrease as seasickness symptoms were alleviated, reinforcing the notion that the BCI facilitates attentional shifting. These findings outline a portable and effective nonpharmacological approach for managing seasickness, along with a mechanistic explanation that aligns with theories of attentional shifting and sensory conflict.

Although several preliminary studies on car sickness suggest that redirecting attention away from the discomfort associated with motion sickness to other objects or tasks may reduce the brain’s awareness of such discomfort \cite{bang2023motion,venkatakrishnan2024,wickens2014structure}, the empirical support for this strategy to alleviate motion sickness is predominantly originates from small–scale, controlled laboratory experiments and lacks thorough validation in real–world scenarios. The considerable challenges associated with implementing attention shifting in practical environments may impede the study of this strategy and its adoption as a widely used method for preventing motion sickness. For example, some attention shifting methods, such as mental arithmetic, require a prominent amount of cognitive load, making the application of such methods difficult to maintain in a state of motion sickness for too long. Another possible reason is that attention shifting can be easily disrupted by the dramatically changing external environment, such as sudden turns, bumps, or other unexpected events on the road, which can quickly draw the passenger’s focus back to the discomfort associated with motion sickness. These factors are particularly pronounced in a maritime setting.  
The duration of maritime travel is also quite long, exposing the brain to prolonged motion sickness conditions that can affect the ability to maintain focus. Furthermore, compared to traveling by car, the rocking motion experienced on a boat is often more pronounced and unpredictable, making attention shifting harder to sustain \cite{kahneman1973attention}. Therefore, currently there is no effective method to prevent seasickness based on attention shifting (or any other nonpharmacological approach).

Among the various methods of redirecting attention, mindfulness meditation is notable for its unique advantages. Unlike techniques that simply shift focus away from distressing thoughts, mindfulness meditation encourages non--judgmental awareness of the present moment, fostering a deep sense of self--understanding and acceptance \cite{zeidan2009mindfulness}. Mindfulness meditation, which involves cultivating an internal focus, enables individuals to discover calmness and stability within themselves; moreover, mindfulness meditation does not rely on external cues or stable environments. These advantages make mindfulness meditation well suited to address the intense motion and physiological challenges associated with seasickness. However, sustained attention can be difficult during mindfulness meditation due to its inherent monotony and lack of intellectual stimulation \cite{nobre2010attention, manly2003enhancing}. Studies have shown that during mindfulness meditation, individuals intermittently monitor themselves to prevent their minds from wandering \cite{coull1998}. If mind wandering does not occur, the individual continues performing mindfulness meditation until spontaneous self--monitoring occurs. When the mind wanders, individuals actively adjust to refocus on the task \cite{davidson2015conceptual}. However, this self--monitoring mechanism may be difficult in nautical environments, making it difficult for individuals to sustain their attention on mindfulness meditation for extended periods without the guidance of an instructor \cite{robertson2010vigilant,montero2021teachers}. 

Among the various methods of redirecting attention, tasks derived from mindfulness meditation are distinguished by its unique advantages. Firstly, tasks performed in a mindfulness-oriented manner require lower cognitive load compared to other attention-demanding tasks such as mental arithmetic \cite{carissoli2015does}. Secondly, mindfulness meditation would enhance the internal regulation of the brain and increase attention engagement more effectively than simply listening to music or breathing \cite{brandmeyer2019neuroscience}. In general, these tasks are characterized by ease of implementation and sustainability, rendering them highly suitable for addressing the intense motion and physiological challenges associated with seasickness. However, sustained attention can be difficult during low cognitive load tasks due to its inherent monotony and lack of intellectual stimulation \cite{manly2003enhancing,nobre2010attention}. Studies have shown that during this kind of tasks, individuals intermittently monitor themselves to prevent their minds from wandering \cite{coull1998neural}. If mind wandering does not occur, the individual continues performing tasks until spontaneous self–monitoring occurs. When the mind wanders, individuals actively adjust to refocus on the task \cite{davidson2015conceptual}. However, this self–monitoring mechanism may be difficult in nautical environments, making it difficult for individuals to sustain their attention on low cognitive load tasks for extended periods without the guidance of an instructor \cite{montero2021teachers,robertson2010vigilant}.

The BCI is an effective self--monitoring system that can provide real--time feedback to participants. It informs individuals of their brain activity, enabling them to develop awareness and, consequently, exercise voluntary control over their self--regulation, ultimately promoting sustained focus during tasks \cite{THIBAULT2016247}. When integrated with task of mindfulness meditation, the proposed BCI converts EEG--based attention scores into audiovisual neurofeedback, serving as an external monitor of the attention state of the individuals. This, in turn, enhances an individual’s self--monitoring capacity and reinforces self--regulation during tasks. Additionally, audiovisual neurofeedback provides participants with additional focal points, facilitating their ability to shift their attention. Our results revealed a reduced theta/beta ratio, a well--established physiological indicator of attention regulation \cite{ogrim2012quantitative,putman2014eeg}, compared to those in both the RS and the PMS. Moreover, participants achieved higher attention scores in the RMS than in both the RS and the PMS. The observed reduction in the theta/beta ratio, coupled with the elevated attention scores, indicates a significant increase in sustained attention during maritime sailing. These findings underscore that the proposed BCI is an effective strategy for attention shifting and that attention shifting can act as a mediator between this intervention and alleviation of seasickness. 

The effectiveness of our system in preventing seasickness has been demonstrated through both behavioral reports and EEG analysis. In terms of behavioral results, most participants (35 out of 43, 81.39\%) reported that the BCI, which offers real--time audiovisual feedback, significantly reduced their seasickness symptoms. Conversely, 69.77\% of the participants in the PMS indicated that their symptoms were not alleviated. Specifically, in real--world maritime sailing experiments, the MISC score increased with time in the RS and PMS groups but was significantly reduced with the application of BCI in the RMS group. The average MISC score in the RMS was significantly lower than in the RS and PMS. When results were stratified by MISC, it appears that the BCI is particularly advantageous for individuals with a serve symptom of seasickness. Upon examination of the behavioral outcomes within the placebo group, these findings indicate an effect that extends beyond a mere placebo phenomenon. Several EEG patterns have been correlated with motion sickness. For example, an increase in spectrum power was observed across all frequency bands with increasing levels of motion sickness \cite{chen2010}. Chen et al., 2010 \cite{chen2010} posited that the augmentation of broadband power might suggest that conflicts within multimodal somatosensory systems cause brain circuits to operate more intensively than when motion sickness is absent, thereby enhancing foundational brain processes. Besides, increased theta relative power \cite{chen2010,krugliakova2020changes,malik2015eeg,wood1994habituation,wood1991effect,wu1992eeg} and beta relative power \cite{naqvi2015,kim2005characteristic} have been proposed as potential EEG signatures of motion sickness in several previous studies conducted in simulated laboratory environments. Our results revealed a decrease in the overall EEG band power in the RMS compared to that in the RS and PMS. The decrease in the overall EEG band power in the RMS consistent with the results mentioned by Chen et al., 2010, who found an increased overall EEG band power in the motion sickness state. Corresponding to the explanation by Chen et al., 2010, our results of decreased overall EEG band power in the RMS indicating that BCI relieve the excessive brain processes. Besides, our research revealed that the prefrontal theta relative power decreased, whereas the prefrontal beta relative power increased during RMS as compared to RS and PMS, indicating that EEG signatures of seasickness mediated by the BCI. It is crucial to note that relative power provides an indication of the actual contribution of a specific band to the overall EEG power. Therefore, the observed increase in prefrontal beta relative power during RMS is not inconsistent with the overall decrease in EEG band power during this condition. These potential EEG signatures of seasickness offering objective proof that underscores the effectiveness of BCI interventions in alleviating seasickness symptoms.


Our results can be well explained by the widely accepted sensory conflict theory for the pathology of seasickness (or any type of motion sickness), which suggests that motion sickness arises from discrepancies between sensory inputs and the expectations of the central nervous system \cite{reason1978motion}. For instance, on a rocking ship, the vestibular system may detect motion, whereas the visual system, if focused on a stable object on the ship, may not perceive any movement. This mismatch can lead to confusion in the brain, triggering symptoms of seasickness such as nausea, dizziness, and disorientation \cite{oman1990motion}. From a neurocognitive model standpoint, the brain can process information via a bottom--up or top--down pathway \cite{knudsen2007fundamental}. During bottom--up processing, external information is passively received by the brain. Conversely, in a top--down approach, the brain actively selects information to process internally. We hypothesis that the sensory conflict may be alleviated by BCI through both external and internal ways. Externally, the proposed BCI employs neurofeedback to provide participants scenes with harmonized audiovisual stimuli, devoid of sensory conflict. These scenes are harmonized because the stimuli within them are generated based on the attention levels of the subjects and are aligned with the subjects' expectations. Since there exists a finite pool of information processing resources and that utilizing capacity for one task limits availability for another \cite{kahneman1973attention}, the synchronization between individual's expectations and mindfulness scenes may partially mask the conflicting from the rocking of the ship and the visually stable cabin. Internally, the BCI facilitates individuals' development of awareness and, consequently, the exertion of voluntary control over their self--regulation, ultimately fostering sustained top--down focus. From the perspective of the neurocognitive model, the regulatory processes of bottom--up and top--down mechanisms are intrinsically competitive with one another \cite{legrain2009neurocognitive}. Individuals can intentionally allocate their attention in a top--down manner to the BCI, thereby modulating the bottom--up allocation of attention concerning sensory conflict information associated with motion sickness. This diminishes an individual's perception of sensory conflict, thereby alleviating symptoms of motion sickness. In conclusion, BCI can alleviate motion sickness through the external delivery of synchronized audiovisual stimuli, thereby masking sensory conflicts, as well as through the internal regulation of attention allocation from bottom--up sensory conflict information to top--down low cognitive load task.

There are several significant future considerations arising from this study that deserve attention. Our primary focus was on the immediate alleviation of seasickness symptoms. Future research should investigate whether the prolonged therapeutic application of this system can lead to a lasting decrease in seasickness susceptibility. Additionally, although this study concentrated on seasickness, the potential of the system to mitigate other forms of motion sickness, such as car sickness, air sickness or visually induced motion sickness, should be further explored.

In conclusion, we introduce an inventive BCI system aimed at relieving seasickness, which was validated through real--world maritime sailing experiments, proving its effectiveness. With the BCI system, participants mainly reported no or mild symptoms. This study underscores that the proposed BCI reduces seasickness by shifting attention. This is the first highly effective, wearable, nonpharmacological solution without side effects for easing seasickness, emphasizing a new application of BCI technology.

\section*{Conflict of interest }
The authors declare that they have no competing interests.

\section*{Acknowledgments}
This work was supported in part by the STI 2030--Major Projects, China under Grant 2022ZD0208900; in part by the Key Research and Development Program of Guangdong Province, China under Grant 2018B030339001; in part by the National Natural Science Foundation of China under Grant No. 62406173; in part by the Guangdong Natural Science Foundation under Grant No. 2024A1515011690.

\section*{Author contributions}
Yuanqing Li was responsible for the conceptualization and supervision of this study. Yuanqing Li, Xiaoyu Bao, Kailin Xu, Qiyun Huang designed and organized the experiments, Yuanqing Li, Xiaoyu Bao, Haiyun Huang and etc. developed the mindfulness-based BCI system. Xiaoyu Bao, Kailin Xu and Jiawei Zhu collected the data. Xiaoyu Bao, Kailin Xu, Jiawei Zhu, Kangning Li and Yuanqing Li jointly analyzed the data. Xiaoyu Bao, Kailin Xu, Jiawei Zhu, Qiyun Huang and Yuanqing Li jointly wrote the paper.

\section*{Data availability}
The data that support the findings of this study are available from the corresponding authors upon reasonable request.

\bibliographystyle{unsrt}  
\bibliography{motion_sickness}

\end{document}